\documentclass[prb,aps,showpacs,twocolumn,unsortedaddress,superscriptaddress]{revtex4}
\usepackage{graphics,bm}
\usepackage{amssymb}
\usepackage{epsfig}
\usepackage{epsf}
\usepackage[usenames]{color}

\begin{document}

\title{Functional Keldysh Theory of Spin Torques}

\author{R.A. Duine}
\email{duine@phys.uu.nl} \homepage{http://www.phys.uu.nl/~duine}

\affiliation{Institute for Theoretical Physics, Utrecht
University, Leuvenlaan 4, 3584 CE Utrecht, The Netherlands}
\date{\today}

\author{A. S. N\'u\~nez}
\email{alvaro.nunez@ucv.cl}
\homepage{http://www.ph.utexas.edu/~alnunez}

\affiliation{ Instituto de F\'isica, PUCV Av. Brasil 2950,
Valpara\'iso, Chile}

\author{Jairo Sinova}
\email{sinova@physics.tamu.edu}
\homepage{http://faculty.physics.tamu.edu/sinova/}

\affiliation{Department of Physics, Texas A\&M University, College
Station, TX 77843-4242, USA}

\author{A.H. MacDonald}
\email{macd@physics.utexas.edu}
\homepage{http://www.ph.utexas.edu/~macdgrp}

\affiliation{The University of Texas at Austin, Department of
Physics, 1 University Station C1600, Austin, TX 78712-0264, USA}

\begin{abstract}
We present a microscopic treatment of current-induced torques and
thermal fluctuations in itinerant ferromagnets based on a
functional formulation of the Keldysh formalism. We find that the
nonequilibrium magnetization dynamics is governed by a stochastic
Landau-Lifschitz-Gilbert equation with spin transfer torques. We
calculate the Gilbert damping parameter $\alpha$ and   the
non-adiabatic spin transfer torque parameter $\beta$ for a model
ferromagnet. We find that $\beta \neq \alpha$, in agreement with
the results obtained using imaginary-time methods of Kohno, Tatara
and Shibata [J. Phys. Soc. Japan {\bf 75}, 113706 (2006)]. We
comment on the relationship between $s-d$ and isotropic-Stoner toy
models of ferromagnetism and more realistic
density-functional-theory models, and on the implications of these
relationships for predictions of the $\beta/\alpha$ ratio which
plays a central role in domain wall motion. Only for a
single-parabolic-band isotropic-Stoner model with an exchange
splitting that is small compared to the Fermi energy does
$\beta/\alpha$ approach one. In addition, our microscopic
formalism incorporates naturally the fluctuations needed in a
nonzero-temperature description of the magnetization. We find that
to first order in the applied electric field, the usual form of
thermal fluctuations via a phenomenological stochastic magnetic
field holds.
\end{abstract}


\maketitle

\def\bx{{\bf x}}
\def\bk{{\bf k}}
\def\half{\frac{1}{2}}
\def\args{(\bx,t)}
\def\bp{{\bf p}}
\def\bq{{\bf q}}

\section{Introduction} \label{sec:intro} Phenomena related to order-parameter manipulation by transport currents have
 recently received a great deal of attention in magnetic
metals and magnetic semiconductors. Spin transfer torques, which
lead to current-driven nanomagnet reversal and to domain wall
motion in narrow wires, have been at the center of this activity
\cite{slonczewski1996,berger1996,bazaliy1998,rossier2004,tsoi1998,myers1999,berger1984,freitas1985,
tatara2004,zhang2004,waintal2004,barnes2005,thiaville2005,rebei2005,ohe2006,xiao2006,tserkovnyak2006,
kohno2006,piechon2006,duine2006,grollier2003,tsoi2003,yamaguchi2004,
klaui2005,beach2006,hayashi2007,yamanouchi2004,yamanouchi2006}. In
spin-transfer-torque theory, the usual Landau-Lifschitz-Gilbert
(LLG) equation of motion for the magnetization direction $\hat
\Omega$ acquires terms corresponding to so-called adiabatic and
non-adiabatic spin transfer torques which are both proportional to
current. Both torques can be constructed from symmetry arguments
by requiring that they be orthogonal to the magnetization
direction and by realizing that the current essentially breaks
inversion symmetry. The latter implies that, in the
long-wavelength limit, terms proportional to $\nabla \hat \Omega$
are allowed in the LLG equation of motion. The adiabatic spin
transfer torque \cite{bazaliy1998,rossier2004} is defined as
$-\left( {\bf v}_{\rm s} \cdot \nabla \right) \hat \Omega$, where
${\bf v}_{\rm s}$ is a velocity, proportional to the current, that
characterizes the efficiency of spin transfer and is required for
dimensional reasons. The non-adiabatic spin transfer torque
\cite{zhang2004} is given by $-\beta \hat \Omega \times \left(
{\bf v}_{\rm s} \cdot \nabla \right) \hat \Omega$ and is
characterized by the dimensionless parameter $\beta$.

The LLG equation that incorporates both spin
transfer torques is then given by
\begin{equation}
\label{eq:LLGwithSTTs} \left( \frac{\partial }{\partial t} + {\bf
v}_{\rm s} \cdot \nabla  \right) \hat \Omega - \hat \Omega \times
{\bf
 H} = - \alpha \hat \Omega \times  \left( \frac{\partial }{\partial
 t}+ \frac{\beta}{\alpha} {\bf
v}_{\rm s} \cdot \nabla \right) \hat \Omega~,
\end{equation}
in the long-wavelength low-frequency limit, \cite{note1} where
${\bf H}$ is the effective field and $\alpha$ the Gilbert damping
constant. The coefficients $\alpha$ and $\beta$ are dissipative,
in the sense that they relate quantities that are even under time
reversal to quantities that are odd under time reversal.
\cite{murakami2004,note2}
\begin{scriptsize}\end{scriptsize} The observation that both
$\alpha$ and $\beta$ are related to dissipation is important in
the context of nonzero-temperature effects which play an
especially important role \cite{duine2006} in experiments with
magnetic semiconductors. \cite{yamanouchi2004,yamanouchi2006}
Nonzero temperature is usually accounted for by adding a Gaussian
stochastic magnetic field ${\bf h}$ to the effective field in the
LLG equation.
\cite{brown1963,kubo1970,ettelaie1984,garciapalacios1998,smith2001,heinonen2004,safonov2005,rossi2005}
The strength of this noisy magnetic field is related to the
Gilbert damping by the fluctuation-dissipation theorem, which
ensures that the system is characterized in equilibrium by the
appropriate Boltzmann distribution. {\it A priori} the
generalization of Eq.~(\ref{eq:LLGwithSTTs}) to nonzero
temperatures is not clear, since both $\alpha$ and $\beta$
correspond to dissipative processes and the
fluctuation-dissipation theorem need not be valid because the
nonzero current implies that the system is out of equilibrium.

As noted in the literature on current-driven domain wall motion,
\cite{barnes2005,tserkovnyak2006} the case of $\beta=\alpha$ is
special because both sides of Eq.~(\ref{eq:LLGwithSTTs}) then
contain the ``co-moving'' derivative $D/Dt =
\partial/\partial t + {\bf v}_{\rm s} \cdot \nabla$, so that the
equation of motion admits solutions $\hat \Omega_{\rm d} (t) =
\hat \Omega_0 (\bx - {\bf v}_{\rm s} t)$ where $\hat \Omega_0
(\bx)$ is a time-independent solution of the LLG equation in the
absence of currents. The solution $\hat \Omega_{\rm d} (t)$
corresponds to ``drift'' of static magnetization textures with
velocity ${\bf v}_{\rm s}$. Arguing that these solutions must
exist Barnes and Maekawa \cite{barnes2005} claim that
$\beta=\alpha$. However, in realistic systems there is no Galilean
invariance \cite{rossier2004} that requires the existence of such
solutions and therefore in general $\beta \neq \alpha$.
\cite{kohno2006,piechon2006} Instead, the non-adiabatic spin
transfer torque acquires contributions from all microscopic
processes that violate spin conservation and therefore correspond
to terms in the microscopic Hamiltonian that are not invariant
under spin rotations. Such processes also contribute to the
Gilbert damping term and therefore in principle any nonzero
Gilbert damping parameter $\alpha$ implies  nonzero non-adiabatic
spin transfer torques, as we show in our specific microscopic
model calculations.

The LLG equation [Eq.~(\ref{eq:LLGwithSTTs})] is motivated mainly
by symmetry considerations and contains four different quantities
whose meaning can be specified precisely only by a microscopic
theory which details, at least in principle, precisely how they
should be evaluated given the full system Hamiltonian. These
quantities are: {\it i}) the effective magnetic field ${\bf H}$,
{\it ii}) the transport spin velocity ${\bf v}_s$, {\it iii}) the
Gilbert damping parameter $\alpha$, and {\it iv}) the
non-adiabatic spin-transfer torque parameter $\beta$.  The
effective magnetic field ${\bf H}$ includes the external magnetic
field and additional contributions due to magnetostatic
interactions and magnetocrystalline anisotropy. The physics of
${\bf H}$ is well understood\cite{aharonibook} and not the subject
of this paper. The three remaining quantities emerge in a
microscopic theory from the slow (up to first order in time
derivatives or frequency $\omega$) smooth (up to first order in
space derivatives or wavevector ${\bf q}$) response of the
magnetization direction to an external magnetic field, in the
presence of an external electric field which drives a transport
current. The coefficient $\alpha$ then emerges as the ratio of the
reactive and dissipative contributions that appear at first order
in $\omega$ in this response function. When spin-orbit
interactions are neglected, it is easy to verify that the
coefficient of the reactive term in the total spin-response is the
unperturbed spin-density, explaining the unit value of this
coefficient in Eq.~(\ref{eq:LLGwithSTTs}). The two first order
space derivative terms in this equation reflect respectively the
change in the reactive and dissipative response due to an external
electric field.  Like its zero-current counterpart, the
current-related reactive terms can be understood in quite general
terms based only on  spin-conservation considerations, while the
dissipative term is sensitive to microscopic details.

As explained above the condition $\beta=\alpha$ corresponds to
Galilean invariance at a macroscopic level. Since the dissipative
terms emerge from spin-dependent disorder (or spin-independent
disorder when spin-orbit interactions are included in the cyrstal
band structure), it is clear that Galilean invariance does not
hold microscopically.  Our calculations show that $\beta \approx
\alpha$ can occur in models with very specific
properties\cite{tserkovnyak2006}, but does not occur in general.
For example $\beta \approx \alpha$ occurs in the specific toy
model that we study below only when the ferromagnetism is weak in
the sense that the exchange splitting is much smaller than the
Fermi energy \cite{tserkovnyak2006}. We believe that we obtain
this result only because the model has isotropic parabolic bands
and that $\beta=\alpha$ (macroscopic Galilean invariance) can
occur only accidentally in systems with either realistic bands or
realistic disorder.  In the important transition metal ferromagnet
spintronic materials in particular, we will argue that orbitals
which have dominant $d$-character contribute more strongly to the
magnetization than to transport and that $\beta$ will tend to be
larger than $\alpha$ as a consequence.

In this paper we present a microscopic derivation of the equation
of motion in Eq.~(\ref{eq:LLGwithSTTs}) of the direction of
magnetization in the presence of current {\it and} at finite
temperature. We use a functional formulation of the Keldysh
non-equilibrium formalism \cite{stoof1999,kamenev2004} which leads
in a natural way to the path-integral formulation of stochastic
differential equations.\cite{zinnjustinbook,duine2002} Within our
microscopic treatment the dissipative nature of $\alpha$ and
$\beta$ is explicit because, as briefly discussed above and to be
shown in more detail, they follow from the dissipative part of the
spin-density spin-density response function and photon two-magnon
interaction vertex, respectively. We focus on the simple
microscopic toy model used in previous work
\cite{tserkovnyak2006,kohno2006}  that is intended to provide a
qualitative description of a generic ferromagnet and includes
disorder and short-range repulsive electron-electron interactions.
The model's ferromagnetism is treated at the level of Stoner
mean-field theory. For the disorder we use the same model as in
Ref.~[\onlinecite{kohno2006}] and, where applicable, our results
for $\alpha$ and $\beta$ agree with theirs. Our random phase
approximation treatment of Stoner quasiparticle fluctuations
evinces the equivalence of Stoner and $s-d$ models, in the sense
that in both models the quantities $\alpha$ and $\beta$ are
determined by the same response function. In particular $\alpha
\neq \beta$ for both models in general.

One benefit of the concinnity of the functional formulation of the
Keldysh formalism is that it enables a natural determination of
the thermal fluctuations without explicitly appealing to the
fluctuation-dissipation theorem. We find that to lowest order in
the applied electric field the form usually assumed for the
strength of the fluctuations holds and that there is no
contribution to the white-noise thermal fluctuations that is
related to the non-adiabatic torque. We emphasize that this
formalism, which has been reviewed in other publications and
applied to other problems,\cite{stoof1999,kamenev2004} is similar
in structure to the functional formulation of standard equilibrium
Green's functions for linear response theory, but is more powerful
for non-equilibrium and non-linear problems.

Since the formalism we use may not be familiar to most readers, we
first present the model and main results in a separate section,
namely Sec.~\ref{sec:modelsummary}. In Sec.~\ref{sec:noneq} we
present the formalism and  outline the calculations. In the
appendix we carry out a typical calculation in more detail. Both
Sec.~\ref{sec:noneq} and the appendix may be skipped by readers
who are familiar with the formalism or who may be more interested
in the results obtained. We end in Sec.~\ref{sec:concl} with our
conclusions.

\section{Model and Summary of Results} \label{sec:modelsummary} We
model the disordered itinerant ferromagnet as electrons with
delta-function-like repulsive interactions, using the Hamiltonian,
\begin{eqnarray}
\label{eq:fullmicrohamiltonian} H[\hat {\bm{\psi}^\dagger}, \hat
{\bm{\psi}}] &=& \int\!d\bx
  \left\{ \hat {\bm{\psi}^\dagger} (\bx,t) \left[
  - \frac{\hbar^2\nabla^2}{2m}
  - \frac{\Delta_{\rm ext}}{2} \tau_z + V_0 (\bx)
  \right. \right. \nonumber \\
  && \left.
 + V_a (\bx) \tau_a \rule{0mm}{5mm} \right]
 \hat {\bm{\psi}} (\bx,t)   + \frac{1}{c} \hat {\bf J} (\bx,t) \cdot {\bf A} (\bx,t) \nonumber \\
  &&  \left. + U
  \hat \psi^\dagger_{\uparrow} (\bx,t) \hat \psi^\dagger_\downarrow (\bx,t)
  \hat \psi_\downarrow (\bx,t) \hat \psi_\uparrow (\bx,t)
 \rule{0mm}{5mm} \right\}~,
\end{eqnarray}
where for notational convenience we have introduced the spinor
\begin{eqnarray}
\label{eq:defspinor}
  \hat {\bm{\psi}} (\bx,t) = \left(
    \begin{array}{c}
      \hat \psi_\uparrow (\bx,t) \\
      \hat \psi_\downarrow (\bx,t)
    \end{array} \right)~.
\end{eqnarray}
In these expressions the Heisenberg operators $\hat \psi_\sigma
(\bx,t)$ annihilate an electron in the spin state labelled by
$\sigma \in \{\uparrow,\downarrow\}$, and obey the usual
equal-time commutation relations. These spin states have their
quantization axis parallel to an external Zeeman magnetic field in
the $z$-direction which contributes $\Delta_{\rm ext}$ to the
energy difference between minority and majority spins. Note that
in Eq.~(\ref{eq:fullmicrohamiltonian}) the Pauli matrices are
indicated by $\tau_a$, and that a sum over the repeated index $a
\in\{x,y,z\}$ is implied. The free-electron dispersion at momentum
$\hbar \bk$, given by $\epsilon_\bk=\hbar^2 \bk^2/2m$, is
parabolic with an effective mass $m$ ($\hbar$ denotes Planck's
constant).

We choose a delta-function interaction with strength $U$ because then the
field-theoretic procedure to introduce the magnetization direction
as a dynamic variable is easier to implement. This so-called
Hubbard-Stratonovich transformation \cite{kleinert1978} can
also be generalized to spatially nonlocal interactions.\cite{stoof2001} This procedure, to be discussed in more detail
in the next section, also yields the mean-field, {\it i.e.},
Stoner, saddle-point equation for the exchange-interaction
contribution to the spin splitting
\begin{eqnarray}
\label{eq:exchangesplitting}
  \Delta &=& U \int \frac{d\bk}{(2\pi)^3}
  \left\{ N_{\rm F} \left[\epsilon_\bk-\frac{(\Delta+\Delta_{\rm ext})}{2}-\mu\right] \right. \nonumber \\
 &&  \left.  -N_{\rm F}
  \left[ \epsilon_\bk+\frac{(\Delta+\Delta_{\rm ext})}{2}-\mu\right] \right\}=U\rho_s~,
\end{eqnarray}
where $N_{\rm F} (x) = \left[ \exp \left( x/k_{\rm B} T
\right)+1\right]^{-1}$ is the Fermi distribution function with
$k_{\rm B} T$ the thermal energy, $\mu$ is the chemical potential
that includes a Hartree mean-field shift, and $\rho_s$ is the
magnetization density. In practice we do not explicitly determine
the exchange splitting from this equation, but simply assume
$\Delta$ is a solution whose value may be determined from
experiment if needed. This is another reason for simply using a
delta function interaction.

For the disorder we use the same model as Kohno {\it et al.}
\cite{kohno2006} in which the spin-dependent disorder potentials
 are characterized by
\begin{equation}
\label{eq:twopointofdisorder} \overline{V_a (\bx) V_{b} (\bx')} =
  \sigma_a \delta (\bx-\bx') \delta_{ab}~,
\end{equation}
where $\overline{\cdots}$ indicates averaging over different
realizations of the disorder. For randomly distributed scatterers
\begin{equation}
\label{eq:impurities}
  \sigma_{x,y} = n_{\rm s} u_{\rm s}^2 \overline{ S^2_{\perp}},~\sigma_z =
  n_{\rm s} u_{\rm s}^2 \overline{S_z^2},~\sigma_0 = n_{\rm i}
 u_{\rm i}^2~,
\end{equation}
where $u_{\rm i}$ ($u_{\rm s}$) and $n_{\rm i}$ ($n_{\rm s}$) are
the strength and density of the scatterer charge (spin) component,
respectively,  and $\overline{S_a^2}$ denotes the average
scatterer field orientation. Within the self-consistent Born
approximation the decay rate $\gamma_\sigma$ and lifetime
$\tau^{\rm sc}_\sigma$ of a plane wave with spin state
$|\sigma\rangle$ are determined from
\begin{equation}
\label{eq:decayrate}
  \hbar  \gamma_\sigma \equiv \frac{\hbar}{2 \tau^{\rm sc}_\sigma}
  = \pi n_{\rm i} u_{\rm i}^2 \nu_\sigma + \pi n_{\rm s} u_{\rm s}^2 \left(
  2 \overline{ S^2_{\perp}} \nu_{-\sigma} + \overline{S_z^2} \nu_\sigma
  \right)~,
\end{equation}
where the density of states per spin at the Fermi level
$\nu_\sigma = m k_{{\rm F}\sigma}/2\pi^2\hbar^2$, and the Fermi
wave number $k_{{\rm F}\sigma} = \sqrt{2m(\epsilon_{\rm F}+\sigma
M)/\hbar^2}$, where $M=(\Delta+\Delta_{\rm ext})/2$ is the total
spin splitting.

Finally, the current in our theory is induced by an external
homogeneous electric field ${\bf E}$ that, in the London gauge, is
related to the vector potential by
\begin{equation}
\label{eq:vectorpotential}
  {\bf A} (t) = \frac{c {\bf E}}{i \omega_{\rm p}} e^{-i\omega_{\rm p}
  t}~,
\end{equation}
where $\omega_{\rm p}$ is the frequency of the electric field, to
be taken to zero eventually, and $c$ is the speed of light. In the
Hamiltonian [Eq.~(\ref{eq:fullmicrohamiltonian})] the vector
potential is minimally coupled to the electrons via the charge
current-density operator
\begin{eqnarray}
\label{eq:current}
 \hat {\bf J} (\bx,t) &=& \frac{i |e| \hbar}{2m} \left[
    \hat { \bm{\psi}}^\dagger (\bx,t) \nabla \hat  {\bm{\psi}} (\bx,t)
  - \left( \nabla \hat {\bm{ \psi}}^\dagger (\bx,t) \right) \hat {\bm{ \psi}} (\bx,t)
  \right]~,
\end{eqnarray}
with $-|e|$ the electron charge. In the above expression we have
omitted the diamagnetic contribution as it plays no role in the
following.

In the next section we derive, starting from the hamiltonian in
Eq.~(\ref{eq:fullmicrohamiltonian}), the equations of motion for
long-wavelength deviations $\delta \bm{\Omega}$ of the
magnetization direction from the collinear ground state, defined
by $\hat \Omega = \hat z + \delta \bm{\Omega}$. We find that these
transverse deviations obey the stochastic equations of motion
\begin{eqnarray}
\label{eq:finaleom}
  && \left( \frac{\partial }{\partial t} + {\bf
v}_{\rm s} \cdot \nabla  \right) \delta \Omega_a
  = \epsilon_{ab}\left[\frac{\Delta_{\rm ext}}{\hbar} \delta \Omega_b \right.
  \nonumber \\
  && ~~~~~~~ \left. + \alpha \left( \frac{\partial }{\partial t} + \frac{\beta}{\alpha}{\bf
v}_{\rm s} \cdot \nabla  \right) \delta \Omega_b  - h_b \right]
  ~,
\end{eqnarray}
where a sum over repeated transverse indices $a,b\in\{x,y\}$ is
implied and $\epsilon_{ab}$ is the two-dimensional Levi-Civita
tensor. The Gilbert damping parameter is given by
\begin{equation}
\label{eq:resultgilbertdampingparameter} \alpha =
\frac{2\pi}{\rho_{\rm s}} \left\{
  n_{\rm s} u_{\rm s}^2
   \left[ \overline{S_\perp^2} \left(\nu_\uparrow^2 + \nu_\downarrow^2 \right)
   +2 \overline{S_z^2} \nu_\uparrow \nu_\downarrow \right]
    \right\}~,
\end{equation}
with the magnetization density $\rho_{\rm s}=\Delta/U$. (For the
$s-d$ model $\rho_s$ corresponds to the carrier spin polarization
density.) The velocity ${\bf v}_{\rm s}$ is related to the
electric field by
\begin{equation}
\label{eq:resultspintransfervelocity}
  {\bf v}_{\rm s} =-\frac{|e|{\bf E}}{m \rho_{\rm s}} \left(n_\uparrow \tau^{\rm sc}_\uparrow - n_\downarrow \tau^{\rm sc}_\downarrow \right)~,
\end{equation}
in terms of the density of majority and minority electrons,
denoted by $n_\uparrow$ and $n_\downarrow$. Using the fact that to
linear order in the electric field the current densities of the
majority and minority electrons are determined from ${\bf
j}_\sigma = n_\sigma |e|^2 \tau^{\rm sc}_\sigma {\bf E}/m$, we
observe that the expression for ${\bf v}_{\rm s}$ reduces to the
usual expression $ {\bf v}_{\rm s} = \left( {\bf j}_\uparrow -
{\bf j}_\downarrow \right)/(-|e| \rho_{\rm s})$. Our result for
the $\beta$-parameter reads
\begin{equation}
\label{eq:resultbeta}
  \beta = \frac{2 \pi n_{\rm s} u_{\rm s}^2}{M}
  \left[
    \frac{n_\uparrow \tau^{\rm sc}_\uparrow \left(\overline{S_z^2} \nu_\downarrow\! +\!\overline{S_\perp^2} \nu_\uparrow \right)
    - n_\downarrow \tau^{\rm sc}_\downarrow \left(\overline{S_z^2} \nu_\uparrow\!+\!\overline{S_\perp^2} \nu_\downarrow \right)}
    {\left(n_\uparrow \tau^{\rm sc}_\uparrow\!-\!n_\downarrow \tau^{\rm sc}_\downarrow \right)}
   \right].
\end{equation}
Notice that, as expected, only spin-dependent scattering
contributes to the non-adiabatic torque parameter $\beta$ and the
Gilbert damping parameter $\alpha$.

In addition, we find that the thermal fluctuations via the
stochastic magnetic field ${\bf h}$ are determined by
\begin{equation}
\label{eq:noisecorrsfinal}
  \langle h_a (\bx,t) h_{b} (\bx',t') \rangle_{\rm noise}
  = \frac{2 \alpha k_{\rm B} T}{\hbar \left( \rho_{\rm
  s}/2\right)}\delta (\bx-\bx') \delta (t-t') \delta_{ab}~,
\end{equation}
where the average is over different realizations of the noise. We
stress that this form for the strength of the fluctuations is
derived explicitly, without appealing to the
fluctuation-dissipation theorem. This is important since it is not
{\it a priori} obvious that in the current-carrying situation the
fluctuation-dissipation theorem holds. The form of the strength of
the fluctuations in Eq.~(\ref{eq:noisecorrsfinal}) is however of
the usual form, {\it i.e.}, it is of the same form as inferred by
the equilibrium fluctuation-dissipation theorem. This result comes
about because shot noise \cite{dejong1997} contributions to the
magnetization noise \cite{rebei2005b,foros2005} enter as
higher-order terms in the applied electric field than the linear
response in electric field considered here.

The linear response result in Eq.~(\ref{eq:finaleom}) is
consistent up to ${\mathcal O}(\delta \bm{\Omega})$  with the
Landau-Lifschitz-Gilbert equation that includes both non-adiabatic
and adiabatic spin transfer torques and thermal fluctuations
\begin{eqnarray}
\label{eq:LLGwithSTTsandnoise} && \left( \frac{\partial }{\partial
t} + {\bf v}_{\rm s} \cdot \nabla  \right) \hat \Omega - \hat
\Omega \times \left( {\bf
 H} +{\bf h}\right) \nonumber \\ &&
 = - \alpha \hat \Omega \times  \left( \frac{\partial }{\partial
 t}+ \frac{\beta}{\alpha} {\bf
v}_{\rm s} \cdot \nabla \right) \hat \Omega~,
\end{eqnarray}
where ${\bf h}$ is a stochastic magnetic field that obeys the
correlations given by Eq.~(\ref{eq:noisecorrsfinal}).

\begin{figure}
\vspace{-0.5cm}
\centerline{\epsfig{figure=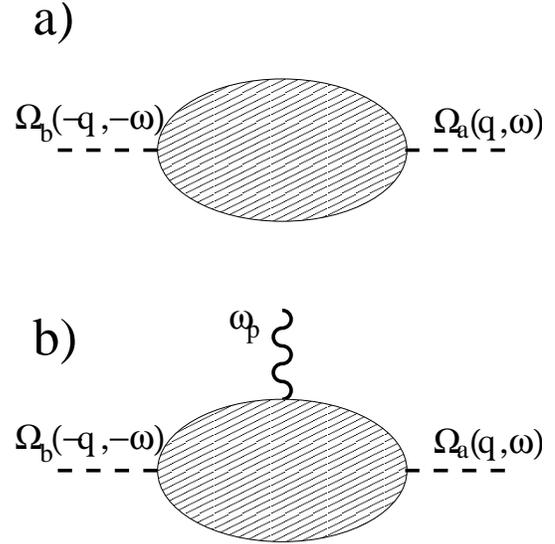, width=7.0cm}}
 \caption{a) Feynman diagram for the transverse spin-density spin-density
 response function. This diagram may equivalently be thought of as
 the spin-wave, or magnon, propagator. The magnon frequency is denoted by $\omega$
 and its momentum by ${\bf q}$. b) Photon two-magnon interaction
 vertex that ultimately gives rises to spin transfer torques. (Note that the photons
 correspond to the external electric field used in our theory to induce transport current
 and hence to terms proportional to the current within linear response.) This vertex
 describes the interaction of spin waves with frequency $\omega$ and momentum ${\bf q}$
 with the electric current that is generated by an external electric (photon) field of
  frequency $\omega_{\rm p}$. (The theory of spin transfer torques is related to the $\omega_{\rm p} \to 0$
  limit of this diagram and this limit is taken in the Feynman diagram.)}
 \label{fig:fullfeynmanresponse}
\end{figure}

We end this section by sketching how the various results come
about. In the theory to be discussed in more detail in the next
section, the two quantities of interest are the transverse
spin-density spin-density response function (or magnon
propagator), which determines the Gilbert damping parameter, and
the photon two-magnon interaction vertex which gives rise to both
the adiabatic and non-adiabatic spin transfer torques. (Note that
the photons simply correspond to the external electric field in
this case.) Feynman diagrams for both of these functions are given
in Fig.~\ref{fig:fullfeynmanresponse}~(a)~and~(b). Quite
generally, the response function in
Fig.~(\ref{fig:fullfeynmanresponse}) has reactive and dissipative
parts. In the long time and length scale expansion corresponding
to the LLG equations, the small-frequency zero-momentum part of
the reactive contribution gives rise to the time-derivative on the
left-hand side of Eq.~(\ref{eq:finaleom}). (Note that after
Fourier transformation frequencies turn into time derivatives.)
The small-frequency zero-momentum part of the dissipative
contribution to the same response function determines the Gilbert
damping term on the right-hand side of Eq.~(\ref{eq:finaleom}).
Physically, this dissipative contribution comes from spin waves
that decay into particle-hole excitations. Energy conservation
then leads a delta-function-like, {\it i.e.}, dissipative,
contribution of the form $\delta (\hbar \omega - \epsilon_1
+\epsilon_2)$ where $\hbar \omega$ is the energy of the spin wave
and $\epsilon_1-\epsilon_2$ the energy of the particle-hole pair.
Summing over all possible particle-hole pair energies and
performing a zero-momentum low-frequency expansion then leads to
the Gilbert damping term on the right-hand side of
Eq.~(\ref{eq:finaleom}). Similarly, the spatial derivatives on the
left-hand side of Eq.~(\ref{eq:finaleom}) are the result of the
reactive contribution to the zero-frequency small-momentum
behavior of the photon spin-wave interaction vertex and give rise
to the adiabatic spin transfer torque. The non-adiabatic torque,
proportional to $\beta$  on the right-hand side of
Eq.~(\ref{eq:finaleom}) then emerges from the dissipative part of
the interaction vertex, and gets contributions from physical
processes in which a spin-wave interacts with the current and
subsequently decays into an incoherent particle-hole excitation.

\section{Nonequilibrium Magnetization Dynamics} \label{sec:noneq}
In this section we derive the stochastic equation of motion for
the transverse magnetization in the presence of current. We start
out by deriving the general equations, and subsequently give the
results for the long-wavelength low-frequency limit. We discuss
the equilibrium situation, {\it i.e.}, the case without electric
field, and the nonequilibrium situation separately.

\subsection{Stochastic Equations of Motion} \label{subsec:equationsofmotion}
Our starting point is the path-integral expression for the
coherent-state probability distribution, written as a functional
integral \cite{stoof1999}
\begin{equation}
\label{eq:fullprobPI}
  P[\bm{\phi}^*,\bm{\phi};t] = \int d[\psi^*_\uparrow] d[\psi_\uparrow] d[\psi^*_\downarrow] d[\psi_\downarrow]
  \exp\left\{ \frac{i}{\hbar}
  S[\bm{\psi}^*,\bm{\psi}]\right\}~.
\end{equation}
Roughly speaking, this distribution specifies the probability for
the system to be in the Grassman coherent state $\bm{\phi}
(\bx,t)$. The action is expressed in terms of the fermionic fields
$\bm{\psi}$ and $\bm{\psi}^*$ by
\begin{eqnarray}
\label{eq:fullaction}
  S[\bm{\psi}^*,\bm{\psi}]&=& \int_{{\mathcal C}^t}\!dt'\int\!d\bx
   \left\{
  \bm{\psi}^*_\sigma (\bx,t')  i \hbar \frac{\partial  \bm{\psi}_\sigma (\bx,t')}{\partial
  t'} \right.  \nonumber \\ && \left.
  -H[\bm{\psi}^*(\bx,t'),\bm{\psi} (\bx,t')]
  \rule{0mm}{5mm}  \right\}~.
\end{eqnarray}
The funtional integration in Eq.~(\ref{eq:fullprobPI}) is over all
fields evolving forward in time from $-\infty$ to $t$, and back,
thereby defining the time integration in the action in
Eq.~(\ref{eq:fullaction}) to be over the Keldysh contour
${{\mathcal C}^t}$.

We rewrite the interaction term as \cite{schulz1990}
\begin{equation}
\label{eq:rewritingint}
 U \psi^*_\uparrow  \psi^*_\downarrow \psi_{\downarrow}
 \psi_{\uparrow}
 = \frac{U}{4} \left(\bm{\psi}^*  \bm{\psi} \right)^2
 - \frac{U}{4} \left( \bm{\psi}^*    \bm{\tau}\cdot \hat n \bm{\psi}
    \right)^2~,
\end{equation}
with $\hat n (\bx,t)$ an arbitrary unit vector that determines the
spin quantization axis. Functional integration over the latter
enforces rotation invariance \cite{schulz1990}. The interaction
terms on the right-hand side of Eq.~(\ref{eq:rewritingint}) are
decoupled by writing them as a Gaussian functional integral over a
density field $\langle\rho (\bx,t) \rangle = \langle\bm{\psi}^*
\bm{\psi} \rangle$, and spin-density field $\langle \Delta (\bx,t)
\hat n (\bx,t) \rangle = U \langle \bm{\psi}^* \bm{\tau}\bm{\psi}
\rangle/2$, respectively. [The precise meaning of the brackets
$\langle \cdots \rangle$ is defined below
Eq.~(\ref{eq:defspindensityrespfunc}).] This Hubbard-Stratonovich
transformation \cite{kleinert1978,stoof1999,schulz1990} then
introduces the density and spin density as dynamical variables in
the path-integral in Eq.~(\ref{eq:fullprobPI}). Density and
spin-density amplitude fluctuations are gapped and can be
approximated at low temperatures and energies by their
saddle-point values. For the density we then find a Hartree-Fock
equation, giving rise to a mean-field Hartree shift which we
absorb in the chemical potential. For the spin-density amplitude
we find the saddle-point equation for $\Delta$ in
Eq.~(\ref{eq:exchangesplitting}).

After these steps we ultimately find that the probability
distribution is given by
\begin{eqnarray}
\label{eq:probdistrmagn}
  P[\bm{\phi}^*,\bm{\phi},\hat \Omega;t] &=& \int d[\psi^*_\uparrow] d[\psi_\uparrow] d[\psi^*_\downarrow] d[\psi_\downarrow]
   d[\hat n] \nonumber \\ &&  \times \exp\left\{\frac{i}{\hbar}
  S'[\bm{\psi}^*,\bm{\psi},\hat n]\right\}~,
\end{eqnarray}
where the unit vector $\hat \Omega$ enters as the boundary
condition at $t'=t$ on the functional integration over the
fluctuating magnetization orientation $\hat n$. We do not
explicitly indicate the boundary condition on the fermion fields,
because, as we shall see, the quantity that enters is the fermion
Green's function which is determined without explicitly referring
to the boundary conditions. The action
$S'[\bm{\psi}^*,\bm{\psi},\hat n]$ is, using the same notation as
for the hamiltonian in Eq.~(\ref{eq:fullmicrohamiltonian}),
explicitly given by
\begin{eqnarray}
\label{eq:actionfullafterHS}
  && S'[\bm{\psi}^*,\bm{\psi},\hat n] = \int_{{\mathcal C}^t}\!dt'\int\!d\bx
  \left\{ \bm{\psi}^* (\bx,t') \left[ i \hbar \frac{\partial}{\partial t'}
  + \frac{\hbar^2\nabla^2}{2m}
  \right. \right. \nonumber \\
  && \left.
  + \frac{\Delta}{2} \hat n (\bx,t') \cdot \bm{\tau}
 + \frac{\Delta_{\rm ext}}{2} \tau_z - V_0 (\bx) -  V_a (\bx) \tau_a  \right]
  \bm{\psi} (\bx,t') \nonumber \\
  && \left.
   - \frac{1}{c} {\bf J} (\bx,t') \cdot {\bf A} (t')\right\}~.
\end{eqnarray}
At this point we note that, if we would add a separate Berry-phase
term in this action to enforce the angular-momentum like
quantization of $\hat n$, the resulting action would be the
starting point for treating the $s-d$ model.

We now do perturbation theory around the collinear state by
writing
\begin{equation}
 \hat n (\bx,t) \simeq \left(
    \begin{array}{c}
      \delta n_x (\bx,t) \\
      \delta n_y (\bx,t) \\
      1- \frac{1}{2} \left( \delta n_x (\bx,t)
 \right)^2 - \frac{1}{2} \left( \delta n_y  (\bx,t) \right)^2    \end{array}
 \right)~,
\end{equation}
and integrate out the electronic fields using second-order
perturbation theory in $\delta n_a (\bx,t)$, and first-order
perturbation theory in ${\bf A} (\bx,t)$. (Note that to find an
equation of motion for $\delta n_a$ that is valid up to
first-order the action needs to be determined up to quadratic
terms in $\delta n_a$.) To this order the effective action for the
magnetization is written in the form
\begin{eqnarray}
\label{eq:effactionfirstever}
 && S_{\rm eff} [\delta {\bf n}] = \int_{{\mathcal C}^t}\!dt'\int\!d\bx
 \left\{
 - \frac{\Delta}{4} \rho_{\rm s} \delta n_a^2 (\bx,t') \right.
 \nonumber
\\
&& +  \int_{{\mathcal C}^t}\!dt''\int\!d\bx'~ \delta n_a (\bx,t') \Pi_{ab} (\bx-\bx';t',t'') \delta n_{b} (\bx',t'') \nonumber \\
 && \left. +  \int_{{\mathcal C}^t}\!dt''\int\!d\bx'~ \delta n_a (\bx,t') K_{ab} (\bx-\bx';t',t'') \delta n_{b} (\bx',t'')
 \right\}, \nonumber \\
\end{eqnarray}
where a sum over repeated transverse indices $a,b\in\{x,y\}$ is
again implied. In Eq.~(\ref{eq:effactionfirstever}), the function
$\Pi_{ab} (\bx;t,t')$ is determined by the transverse spin-density
spin-density response function
\begin{eqnarray}
\label{eq:defspindensityrespfunc}
  && \Pi_{ab} (\bx-\bx';t,t') = \nonumber \\
  &&  \frac{i
\Delta^2}{8\hbar} \langle \bm{\psi}^\dagger (\bx,t) \tau_a
\bm{\psi} (\bx,t) \bm{\psi}^\dagger (\bx',t') \tau_{b} \bm{\psi}
(\bx',t') \rangle~,
\end{eqnarray}
shown diagrammatically in Fig.~\ref{fig:fullfeynmanresponse}~(a).
In this expression the brackets $\langle \cdots \rangle \equiv
{\rm Tr} \left[ \hat \rho (-\infty) \cdots \right]$ denote an
average with respect to the density matrix $\hat \rho (-\infty)$
of a system of electrons with the action in
Eq.~(\ref{eq:actionfullafterHS}), with $\hat n (\bx,t) = \hat z$
and ${\bf A} =0$, that is in equilibrium. The function $K_{ab}
(\bx;t,t')$ is given by
\begin{eqnarray}
\label{eq:defKfunction}
  && K_{ab}(\bx-\bx';t,t') = \int_{{\mathcal C}^t}\!dt''\int\!d\bx''
\left[ \frac{\Delta^2}{8 \hbar^2 c} \right. \nonumber \\
&&\left. \times  \bm{\Lambda}_{ab} (\bx,\bx',\bx'';t,t',t'') \cdot
{\bf A} (t'') \rule{0mm}{5mm} \right]~,
\end{eqnarray}
where the photon two-magnon vertex function
\begin{eqnarray}
&& \bm{\Lambda}_{ab} (\bx,\bx',\bx'';t,t',t'') \nonumber \\
&& = \langle
  \bm{\psi}^\dagger (\bx,t') \tau_a \bm{\psi} (\bx,t')
  \bm{\psi}^\dagger (\bx',t') \tau_{b} \bm{\psi} (\bx',t')
  {\bf J} (\bx'',t'')
 \rangle~, \nonumber  \\
\end{eqnarray}
shown as a Feynman diagram in
Fig.~\ref{fig:fullfeynmanresponse}~(b).

We note at this stage, that the above procedure, i.e., integrating
out the fermionic degrees of freedom after expanding the
Hubbard-Stratonovich fields using second-order perturbation
theory, recovers the random phase approximation (RPA). The usual
structure of the RPA response function contains the Stoner
enhancement factor $1/(\Pi-U)$ where $\Pi$ is the zeroth-order
``bubble" diagram. This form applies to gapped fields such as the
density-density and longitudinal spin-density spin-density
response functions, whose fluctuations we have neglected. The
transverse spin-density spin-density response function does not
have this Stoner enhancement factor.

Next, we split the magnetization into semiclassical and
fluctuating parts, according to
\begin{equation} \label{eq:splitfields}
  \delta n_a (\bx,t_\pm) = \delta \Omega_a (\bx,t) \pm \frac{\xi_a
  (\bx,t)}{2}~,
\end{equation}
where $t_+$ and $t_-$ refer to the forward and backward branches
of the Keldysh contour, respectively. This transformation results
in the action
\begin{widetext}
\begin{eqnarray}
\label{eq:effactionrealaxis}
 && S_{\rm eff} [\delta \bm{\Omega},\bm{\xi}] = \int_{-\infty}^t\!dt'\int\!d\bx
 \left\{
 - \frac{\Delta \rho_{\rm s}}{2}  \delta \Omega_a (\bx,t') \xi_a (\bx,t') \right\}
 \nonumber \\
&& +
\int_{-\infty}^t\!dt'\int\!d\bx\int_{-\infty}^t\!dt''\int\!d\bx'
\left\{ \delta \Omega_a (\bx,t') \left[
 \Pi^{(-)}_{ab} (\bx-\bx';t'-t'') + K^{(-)}_{ab} (\bx-\bx';t'-t'') \right]\xi_{b} (\bx',t'')
 \rule{0mm}{5mm} \right\} \nonumber \\
 && +  \int_{-\infty}^t\!dt'\int\!d\bx\int_{-\infty}^t\!dt''\int\!d\bx' \left\{ \xi_a
(\bx,t') \left[
 \Pi^{(+)}_{ab} (\bx-\bx';t'-t'') + K^{(+)}_{ab} (\bx-\bx';t'-t'')\right] \delta \Omega_{b} (\bx',t'')
 \rule{0mm}{5mm}  \right\} \nonumber \\
 && +  \int_{-\infty}^t\!dt'\int\!d\bx\int_{-\infty}^t\!dt''\int\!d\bx' \left\{2 \xi_a
(\bx,t') \left[
 \Pi^{\rm K}_{ab} (\bx-\bx';t'-t'') +K^{\rm K}_{ab} (\bx-\bx';t'-t'') \rule{0mm}{4mm}  \right] \xi_{b} (\bx',t'')
 \rule{0mm}{5mm}  \right\}~,
\end{eqnarray}
\end{widetext}
where the time integrations are now over the real axis from minus
infinity to $t$.

Before we proceed, we make some general statements about dealing
with functions on the Keldysh contour.\cite{rammer1986} A general
function $A(t,t')$, with time arguments on the Keldysh contour,
can be decomposed into its analytic pieces by means of
\begin{equation}
\label{eq:decomgrlsr}
 A(t,t') \equiv \theta (t,t') A^> (t,t') + \theta (t',t) A^< (t,t')~,
\end{equation}
with $\theta (t,t')$ the Heaviside step function on the Keldysh
contour. Generally there can also be a piece $A^{\delta} \delta
(t,t')$, but such a general decomposition is not needed here.
Retarded and advanced functions, distinguished by the superscripts
$(+)$ and $(-)$, respectively, are related to the analytic pieces
by
\begin{equation}
\label{eq:defretadv}
  A^{(\pm)}(t,t') \equiv \pm  \theta (\pm (t-t') ) \left[ A^> (t,t') - A^< (t,t')
  \right]~.
\end{equation}
In addition, the Keldysh part, which, as we shall see, determines
the strength of the fluctuations, is defined by
\begin{equation}
\label{eq:defkeldyshpart}
  A^{\rm K}(t,t') \equiv  \left[ A^> (t,t') + A^< (t,t')
  \right]~.
\end{equation}
Note that in the effective action
[Eq.~(\ref{eq:effactionrealaxis})] the retarded, advanced, and
Keldysh parts of the various functions depend only on the
difference of time arguments (we have implicitly taken the limit
$\omega_{\rm p} \to 0$).

To derive the equation of motion that is obeyed by the
magnetization $\delta \Omega_a$ we perform another
Hubbard-Stratonovich transformation, and write the part of the
action that is quadratic in the fluctuations $\xi_a$ as a Gaussian
functional integral over an auxiliary field $\eta_a$ which will
turn out to correspond, up to prefactors, to the stochastic
magnetic field {\bf h}. Explicitly, we then have for the
probability distribution that
\begin{eqnarray}
\label{eq:probdistrmagnwithnoise}
  P[\bm{\phi}^*,\bm{\phi},\hat \Omega;t] &=& \int
   d[\delta \bm{\Omega}] d[\bm{\xi}] d[\bm{\eta}] \nonumber \\ &&  \times
   P[\bm{\eta}]\exp\left\{\frac{i}{\hbar}
  S_{\rm eff} [\delta\bm{\Omega},\bm{\xi},\bm{\eta}]\right\}~,
\end{eqnarray}
with the effective action
\begin{widetext}
\begin{eqnarray}
\label{eq:effactionrealaxiswithnoise}
 && S_{\rm eff} [\delta \bm{\Omega},\bm{\xi},\bm{\eta}] = \int_{-\infty}^t\!dt'\int\!d\bx
 \left\{
 - \frac{\Delta \rho_{\rm s}}{2}  \delta \Omega_a (\bx,t') \xi_a (\bx,t') + \eta_a (\bx,t) \xi_a (\bx,t) \right\}
 \nonumber \\
&& +
\int_{-\infty}^t\!dt'\int\!d\bx\int_{-\infty}^t\!dt''\int\!d\bx'
\left\{ \delta \Omega_a (\bx,t') \left[
 \Pi^{(-)}_{ab} (\bx-\bx';t'-t'') + K^{(-)}_{ab} (\bx-\bx';t'-t'') \right]\xi_{b} (\bx',t'')
 \rule{0mm}{5mm} \right\} \nonumber \\
 && +  \int_{-\infty}^t\!dt'\int\!d\bx\int_{-\infty}^t\!dt''\int\!d\bx' \left\{ \xi_a
(\bx,t') \left[
 \Pi^{(+)}_{ab} (\bx-\bx';t'-t'') + K^{(+)}_{ab} (\bx-\bx';t'-t'')\right] \delta \Omega_{b} (\bx',t'')
 \rule{0mm}{5mm}  \right\}~.
\end{eqnarray}
\end{widetext}
This action is now linear in the fluctuations $\bm{\xi}$, and the
funtional integration over these fluctuations leads to a
constraint that is precisely the equation of motion for the
magnetization $\delta \bm{\Omega}$. We find that
\begin{widetext}
\begin{eqnarray}
\label{eq:stocheomfull} && -\left[ \Pi^{(+)}_{ab} \left(-i\nabla,i
\frac{\partial}{\partial t}\right)
 +
 \Pi^{(-)}_{ba} \left(i \nabla,-i \frac{\partial}{\partial
t}\right) + K^{(+)}_{ab} \left(-i\nabla,i \frac{\partial}{\partial
t}\right)
 +
 K^{(-)}_{ba} \left(i \nabla,-i \frac{\partial}{\partial
t}\right) \right]
 \delta \Omega_{b} (\bx,t) \nonumber \\
 && ~~~~~~~~~~~~~~~~~~ + \frac{\Delta \rho_{\rm s}}{2} \delta
 \Omega_a
 (\bx,t)= \eta_a (\bx,t)~,
\end{eqnarray}
\end{widetext}
with $\Pi_{ab} (\bq,\omega)$ and $K_{ab} (\bq,\omega)$ denoting
the Fourier transforms of $\Pi_{ab} (\bx-\bx';t-t')$ and $K_{ab}
(\bx-\bx';t-t')$, respectively. Note that the procedure used in
Eq.~(\ref{eq:splitfields})~and~(\ref{eq:probdistrmagnwithnoise}),
which leads ultimately to the above equation of motion,
circumvents the usual difficulties of deriving an equation of
motion with dissipative terms from an action. The probability
distribution for the noise is given by
\begin{widetext}
\begin{equation}
\label{eq:probdistrnoise}
  P[\bm{\eta}] = \exp \left\{ \frac{i}{\hbar} \int_{-\infty}^t\!dt'\int\!d\bx\int_{-\infty}^t\!dt''\int\!d\bx' \left\{2 \eta_a
(\bx,t') \left[
 \Pi^{\rm K}_{ab} (\bx-\bx';t'-t'') +K^{\rm K}_{ab} (\bx-\bx';t'-t'') \rule{0mm}{4mm}  \right]^{-1} \eta_{b} (\bx',t'')
 \rule{0mm}{5mm}  \right\} \right\}~,
\end{equation}
\end{widetext}
so that the correlation function of the stochastic magnetic field
follows as
\begin{eqnarray}
\label{eq:noisecorrsfull}&& \langle \eta_a (\bx,t) \eta_{b}
(\bx',t')\rangle_{\rm noise}
 = \frac{\hbar}{i} \left[\Pi^{\rm K}_{ab}
 (\bx-\bx';t-t') \right. \nonumber \\
 && \left. +K^{\rm K}_{ab}
 (\bx-\bx';t-t')\right]~.
\end{eqnarray}
The results in
Eqs.~(\ref{eq:stocheomfull})~and~(\ref{eq:noisecorrsfull}) are the
main results of this subsection. Clearly, our main tasks are now
to determine the long-wavelength low-frequency behavior of the
non-equilibrium spin-density spin-density response function and
the photon two-magnon vertex function. These calculations will be
outlined in the next two subsections. We start out with the
equilibrium situation in which the electric field is zero and we
only need to consider the spin-density spin-density response
function.

\begin{figure}
\vspace{-0.5cm} \centerline{\epsfig{figure=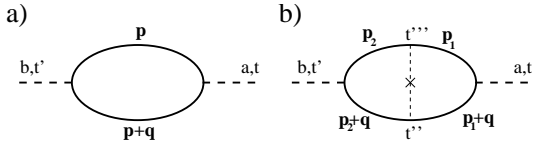,
width=7.0cm}}
 \caption{a) Lowest-order diagram, and b) first vertex
 correction to the spin-density spin-density response function.
 The momentum of the spin wave is denoted by ${\bf q}$.
 Impurity scattering is indicated by the thin dashed line where the $\times$
 indicates the position of the
 impurity.}
 \label{fig:spindensity}
\end{figure}

\subsection{Equilibrium Situation} \label{subsec:equilibrium}
As noted by Kohno {\it et al.} \cite{kohno2006}, because the life
time of the electrons is almost always extremely small compared to
the spin splitting in metallic ferromagnets, {\it i.e.}, $\hbar
\gamma_\sigma \ll M$, we only need to consider the first vertex
correction to the spin-density spin-density response function. The
diagrams that contribute to leading order are given in
Fig.~\ref{fig:spindensity}. The corresponding expression reads
\begin{widetext}
\begin{eqnarray}
\label{eq:respfctzerothplusvertex} && \Pi_{ab} (\bq;t,t') =
\frac{i \Delta^2}{8 \hbar}\left\{ \rule{0mm}{7mm} \int \frac{d
{\bf p}}{(2\pi)^3} {\rm Tr} \left[ \tau_a G ({\bf p}+{\bf q};t,t')
\tau_{b} G({\bf p};t',t)
  \right] \right. \nonumber \\
&&  \left. +\frac{1}{\hbar^2} \sum_{a' \in
  \{0,x,y,z\}}\!\!\!\sigma_{a'} \int\! \frac{d {\bf p}_1}{(2\pi)^3}
  \int\! \frac{d {\bf p}_2}{(2\pi)^3}
  \int_{{\mathcal C}^\infty}\!\!\!dt''\!\!\!\int_{{\mathcal
  C}^\infty}\!\!\!dt'''
  {\rm Tr} \left[ \tau_a G (\bp_1+\bq;t,t'') \tau_{a'}
  G (\bp_2+\bq;t'',t')\tau_{b} G (\bp_2;t',t''')\tau_{a'} G (\bp_1;t''',t)
  \right]
   \right\}~, \nonumber \\
\end{eqnarray}
\end{widetext}
where the trace is over spin space and $\tau_0$ denotes the
$2\times 2$ identity matrix. The first term in this equation
corresponds to the lowest-order diagram in
Fig.~\ref{fig:spindensity}~a) and the second term to the diagram
with the vertex correction in Fig.~\ref{fig:spindensity}~b).

The Green's function is defined as
\begin{eqnarray}
\label{eq:gfkeldysh}
 &&i G_{\sigma\sigma'} (\bx-\bx';t,t') \equiv \langle \psi_\sigma (\bx,t) \psi^*_{\sigma'} (\bx',t')
 \rangle  \nonumber \\
 && = \theta (t,t') {\rm Tr} \left[ \hat \rho (-\infty) \hat \psi_\sigma (\bx,t) \hat \psi^\dagger_{\sigma'} (\bx',t') \right]
 \nonumber \\ &&
 -\theta (t',t) {\rm Tr} \left[ \hat \rho (-\infty)\hat  \psi^\dagger_{\sigma'} (\bx',t') \hat \psi_\sigma (\bx,t)
 \right]~,
\end{eqnarray}
so that the Fourier transforms of its analytic pieces read
\begin{eqnarray}
 -i G^< (\bk, \omega) &=&  A (\bk, \omega) N_{\rm F} (\hbar \omega -
 \mu)~; \nonumber \\
 i G^> (\bk, \omega) &=& A (\bk, \omega) \left[1- N_{\rm F} (\hbar \omega -
 \mu) \right]~,
\end{eqnarray}
where the spectral function $A(\bk,\omega)$ is defined by
\begin{equation}
\label{eq:spectrfct}
  A (\bk,\omega) = i \left[ G^{(+)} (\bk,\omega) - G^{(-)} (\bk,\omega)
  \right]~.
\end{equation}
Finally, the retarded and advanced Green's functions are given by
\begin{equation}
\label{eq:gfretadv}
 G^{(\pm)}_{\sigma\sigma'} (\bk,\omega) = \frac{\delta_{\sigma\sigma'}}
 {\hbar \omega^\pm - \epsilon_\bk   + M \sigma  \pm i \hbar
 \gamma_\sigma}~,
\end{equation}
where $\hbar \omega^\pm = \hbar \omega \pm i 0$ as usual .

With these ingredients, the calculation of the retarded, advanced,
and Keldysh components of the response function in
Eq.~(\ref{eq:respfctzerothplusvertex}), is, in principle,
straightforward. Some details of these calculations are described
in the appendix. Here we directly present the results. For the
retarded and advanced components we find that
\begin{eqnarray}
\label{eq:retadvresponsfctfinalresult}
  &&  \Pi^{(\pm)}_{xx} (\bq,\omega)= \Pi^{(\pm)}_{yy} (\bq,\omega)
   = \nonumber \\ && \frac{\Delta^2 \rho_{\rm s}}{8 M} \pm i \pi \Delta^2 \hbar \omega \left\{
  \frac{ n_{\rm s} u_{\rm s}^2
   \left[ \overline{S_\perp^2} \left(\nu_\uparrow^2 + \nu_\downarrow^2 \right)
   +2 \overline{S_z^2} \nu_\uparrow \nu_\downarrow \right]
   }{8 M^2} \right\}~, \nonumber \\
   &&  \Pi^{(\pm)}_{xy} (\bq,\omega) =-  \Pi^{(\pm)}_{yx} (\bq,\omega)
    = \frac{\Delta^2 \rho_{\rm s}}{16 M^2} i \hbar \omega~,
\end{eqnarray}
with the Keldysh parts given by
\begin{eqnarray}
\label{eq:keldyshtotallowenergy}
 &&  \Pi^{{\rm K}}_{xx} (\bq ,\omega)  =  \Pi^{{\rm K}}_{yy} (\bq ,\omega)
    = \nonumber \\
    &&  i \pi \Delta^2 k_{\rm B} T  \left\{
   \frac{n_{\rm s} u_{\rm s}^2
   \left[ \overline{S_\perp^2} \left(\nu_\uparrow^2 + \nu_\downarrow^2 \right)
   + 2 \overline{S_z^2} \nu_\uparrow \nu_\downarrow \right]
   }{2 M^2}\right\}~, \nonumber \\
&&  \Pi^{{\rm K}}_{xy} (\bq ,\omega)  =  \Pi^{{\rm K}}_{yx} (\bq
 ,\omega)=0~.
\end{eqnarray}
In order to obtain the Gilbert damping coefficient it is
sufficient to perform a zero-momentum small-frequency expansion of
this response function. The first term that enters in a
long-wavelength expansion is quadratic and determines the spin
stiffness that is not of interest to use here. In addition, in
order to determine the fluctuations it turns out to be sufficient
to obtain the zero-momentum zero-frequency part of the Keldysh
response function. Inserting these results into the full equations
of motion in
Eqs.~(\ref{eq:stocheomfull})~and~(\ref{eq:noisecorrsfull})
straightforwardly leads to the results in
Eqs.~(\ref{eq:finaleom}-\ref{eq:noisecorrsfinal}), with ${\bf
v}_{\rm s}=0$. (In arriving at these final results we have taken
the limit $\Delta_{\rm ext} \ll \Delta$.) In the next section we
consider the situation with an external electric field which leads
to a nonzero spin-transfer velocity ${\bf v}_{\rm s}$. We end this
subsection by noting that, from a phenomenological viewpoint,
Eqs.~(\ref{eq:finaleom})~and~(\ref{eq:noisecorrsfinal}) are under
debate \cite{smith2001,safonov2005}, even for ${\bf v}_{\rm s}=0$.
We hope that the microscopic derivation presented here sheds new
light on this controversy. Finally, we note that the temporal
delta function in Eq.~(\ref{eq:noisecorrsfinal}) arises by taking
the zero-frequency limit of the Keldysh part of the spin-density
spin-density response function. This implies that the stochastic
magnetic field in Eq.~(\ref{eq:noisecorrsfinal}) corresponds to a
Stratonovich stochastic process, rather an Ito one
\cite{nicobook}.

\begin{widetext}
\begin{center}
\begin{figure}
\vspace{-0.5cm} \centerline{\epsfig{figure=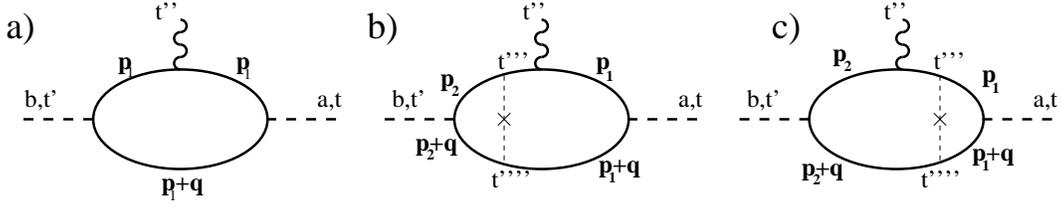,
width=14.0cm}}
 \caption{Feynman diagrams that contribute to the spin-wave current
 interaction that gives rise to spin transfer torques. a) Lowest-order diagram, and b) and c)
 leading-order vertex corrections.}
 \label{fig:intvertex}
\end{figure}
\end{center}
\end{widetext}

\subsection{With Current} \label{subsec:current} Our next task is
to evaluate the function $K_{ab} (\bx;t,t')$ that is proportional
to the photon two-magnon interaction vertex and hence, from a
microscopic point of view, ultimately gives rise to spin transfer
torques. The relevant Feynman diagrams are given in
Fig.~\ref{fig:intvertex} and correspond to the expression
\begin{widetext}
\begin{eqnarray}
\label{eq:Kfuncfull} && K_{ab} (\bq;t,t') = \frac{|e| \Delta^2}{4
 m \hbar}
 \int_{{\mathcal C}^\infty}\!\!dt''\!\int \frac{d \bp_1}{(2
 \pi)^3} \frac{e^{-i \omega_{\rm p} t''}}{\omega_{\rm p}} \left\{
 \rule{0mm}{6mm}
 {\rm Tr}  \left[ \tau_a G (\bp_1+\bq;t,t') \tau_b G(\bp_1;t',t'') G(\bp_1;t'',t)
 \right]\left( \bp_1 \cdot {\bf E} \right) \right. \nonumber \\
 && +\frac{1}{\hbar^2} \sum_{a'\in\{0,x,y,z\}} \sigma_{a'} \int_{{\mathcal C}^\infty}\!\!dt''' \int_{{\mathcal C}^\infty}\!\!dt''''\!
 \int \frac{d \bp_2}{(2 \pi)^3} \left\{\rule{0mm}{5mm} \right. \nonumber \\
 && ~~~~~~~~~~~~\times  {\rm Tr} \left[ \tau_a G (\bp_1\!+\!\bq;t,t'''')\tau_{a'} G(\bp_2\!+\!\bq;t'''',t')
 \tau_b G(\bp_2;t',t''') \tau_{a'} G(\bp_1;t''',t'') G(\bp_1;t'',t)
 \left( \bp_1 \cdot {\bf E} \right) \right.  \nonumber \\
 && ~~~~~~~~~~ +
\left.\left.  \left. \tau_a G (\bp_1\!+\!\bq;t,t'''')\tau_{a'}
G(\bp_2\!+\!\bq;t'''',t')\tau_b G(\bp_2;t',t'') \tau_{a'}
G(\bp_2;t'',t''') G(\bp_1;t''',t) \left( \bp_2 \cdot {\bf E}
\right)  \right] \rule{0mm}{5mm} \right\}\rule{0mm}{6mm}
\right\}~,
\end{eqnarray}
\end{widetext}
where the trace is again over spin space. In this expression the
first, second and third terms correspond to the Feynman diagrams
in Fig.~\ref{fig:intvertex}~a),~b)~and~c), respectively.
Determining the low-frequency long-wavelength behavior of the
retarded, advanced, and Keldysh components from
Eq.~(\ref{eq:Kfuncfull}) is straightforward but rather tedious.
Typical steps in the calculations are illustrated in the appendix
for the spin-density spin-density response function. Here, we
directly present the results. Note that to obtain the spin
transfer torques, and in particular the $\beta$-coefficient that
characterizes the non-adiabatic spin transfer torque, it is
sufficient to perform a zero-frequency long-wavelength expansion.

The results for the various parts of the function $K_{ab}
(\bq;t,t')$, that ultimately determine the adiabatic spin transfer
torque, are given by
\begin{eqnarray}
\label{eq:finalresultKoffd}
  && K^{\rm K}_{xy} (\bq,\omega) = K^{\rm K}_{yx} (\bq,\omega) =
  0~; \nonumber \\
  && K^{(\pm)}_{xy} (\bq,\omega) =- K_{yx}^{(\pm)} (\bq,\omega) \nonumber \\
  && =
  \frac{i\Delta^2 \hbar |e|\left( \bq \cdot {\bf E} \right)}{16 m M^2}
  \left( n_\uparrow \tau^{\rm sc}_\uparrow - n_\downarrow \tau^{\rm sc}_\downarrow
  \right)~.
\end{eqnarray}
We note that these off-diagonal parts correspond to the reactive
part of the photon two-magnon interaction vertex. The dissipative
part that gets contributions from decay processes and determines
the non-adiabatic spin torque is given by
\begin{eqnarray}
\label{eq:finalresultKdiag}
  && K_{xx}^{(\pm)} (\bq,\omega) = K_{yy}^{(\pm)} (\bq,\omega)
  \nonumber \\ && = \pm \frac{i \Delta^2 \hbar^2 |e| \left( \bq \cdot {\bf E} \right)}{16 m M^3}
  \left[ \rule{0mm}{4mm} \left( n_\uparrow \tau^{\rm sc}_\uparrow \gamma_\downarrow - n_\downarrow \tau^{\rm sc}_\downarrow \gamma_\uparrow
  \right)\right.
  \nonumber \\ && \left. - \frac{\pi}{\hbar} \left( n_{\rm i} u^2 - n_{\rm s} u_{\rm s}^2 \overline{S_z^2} \right)
  \left( n_\uparrow \tau^{\rm sc}_\uparrow \nu_\downarrow - n_\downarrow \tau^{\rm sc}_\downarrow \nu_\uparrow \right)
  \right]~; \nonumber \\
  && K^{\rm K}_{xx} (\bq,\omega) = K^{\rm K}_{yy} (\bq,\omega) =
  0~.
\end{eqnarray}
From the above expression we note that, to first order in the
electric field, the Keldysh part of the photon two-magnon
interaction vertex is zero. Ultimately, this implies that the
current does not alter the thermal fluctuations, at least to first
order perturbation theory in the electric field. Finally, we
remark that inserting the above results from
Eqs.~(\ref{eq:finalresultKoffd})~and~(\ref{eq:finalresultKdiag})
into the general equation of motion in Eq.~(\ref{eq:stocheomfull})
leads in a straightforward manner to the results for ${\bf v}_{\rm
s}$ and $\beta$ presented in Sec.~\ref{sec:modelsummary}.

\section{Conclusions} \label{sec:concl} In conclusion, we have
presented a general framework for the derivation of the effective
equations of motion for the magnetization direction of a metallic
ferromagnet, including nonzero-temperature effects and current. An
important aspect of our approach is that the functional Keldysh
methods we employ enable us to incorporate thermal fluctuations
via stochastic forces in a unifying manner, without explicitly
invoking the fluctuation-dissipation theorem. As a specific
example, we have carried out detailed calculations for the model
of a disordered itinerant ferromagnet used by Kohno {\it et al.}
\cite{kohno2006}. Our results for the Gilbert damping parameter
and the $\beta$-parameter that characterizes the non-adiabatic
torque are identical to the results found by these authors using
imaginary-time methods. We have in addition determined the thermal
fluctuations and found that, although the non-adiabatic torque
corresponds to a dissipative process and the current-carrying
situation makes application of the fluctuation-dissipation theorem
questionable, to first order in the electric field the usual
fluctuation-dissipation relation to the Gilbert damping holds.

The method presented here is quite general, and in the near future
we intend to apply it also to other models of ferromagnets. We
briefly comment on the generality of our results and what to
expect for other models. In practice ferromagnetism in metals is
usually described in terms of some combination of ground state and
time-dependent spin-density-functional (SDF) theory. The structure
of the ground state theory is then the same as that of the
saddle-point mean-field equations that arise in our theory, with
the spin-dependent interaction in our theory replaced by the
spin-dependence of the exchange-correlation potential in SDF
theory and our parabolic bands replaced by more complex bands
specific to a particular material.  For transition metals the more
realistic bands of SDF theory are hybridized $s$ and $d$ bands
with $\bf{k}$ dependent spin-splitting which tends to be larger in
bands with dominant $d$-character.  Transition metal ferromagnets
are sometimes described by a crude model in which hybridization is
not explicitly accounted for and the $d$-orbitals are assumed to
be fully spin-polarized.  In this $s-d$ model the $d$-orbitals do
not contribute to the density of states at the Fermi level since
the majority spins are fully occupied and the minority spins are
empty.  It follows that the $d$-orbitals {\rm do} not contribute
to transport or to any other property that involves only orbitals
at the Fermi energy.  When the formalism of our paper is applied
to an $s-d$ model rather than to the single-band model we discuss,
the $d$ orbitals can contribute to properties associated with the
reactive pieces of the response functions we evaluate, but not to
the properties that come from the low-energy limits of the
dissipative response function pieces.  The $d$-orbitals do
contribute to the coefficient of $\omega$ which translates into
the LLG precessional dynamics time derivative for example.  [See
the last line of Eq.~(\ref{eq:retadvresponsfctfinalresult}). This
$\bf{q}=0$ time-derivative can be interpreted as capturing the
Berry phase associated with adiabatic
spin-dynamics.\cite{auerbachbook}] It follows that the $d$ and $s$
orbital contributions to this coefficient are proportional to
their respective contributions to the total spin-density.  The
$d$-orbitals of an $s-d$ model do {\em not}, however, contribute
to the reactive adiabatic spin-torque (${\bf v}_{\rm s}$) term
because the $d$-bands are either full or empty and therefore do
not respond to an electric field.

The $\alpha$ and $\beta$ dissipative parameters are both defined
as dimensionless ratios of coefficient contributions from
dissipative and non-dissipative terms
 in the equation
of motion Eq.~(\ref{eq:retadvresponsfctfinalresult}).
 Because $\alpha$
parameterizes the ratio of the two time-derivative terms it is
indirectly altered by the $d$-bands. In contrast, $\beta$
parameterizes the ratio of the two space-derivative terms neither
of which has a $d$-orbital contribution.  This is the reason, as
noted in previous
studies,\cite{zhang2004,tserkovnyak2006,kohno2006} why $\beta$
tends to be larger than $\alpha$ in $s-d$ models, especially when
the $d$ orbitals make the dominant contribution to the
spin-density.  As we have mentioned previously, and originally
shown by Tserkovnyak {\it et. al.} \cite{tserkovnyak2006}, for
spin-dependent scattering models with parabolic dispersion
$\alpha\approx\beta$ in a Stoner band model when the exchange
splitting is much smaller than the Fermi energy, but not when the
$d$-orbital Berry phase is added to the reactive time derivative
of an $s-d$ model. It is perhaps expected that $\alpha$ should
approximately equal to $\beta$ in this limit since all the
ingredients necessary for macroscopic Galilean invariance seem to
be present. When the spin-polarization is small there is little to
distinguish one direction of spin-polarization from another and
therefore one position in a spin texture from another. When the
bands are parabolic in addition, an external electric field simply
accelerates the system's center of mass. Explicit calculations for
the present toy model demonstrate conclusively that the $\alpha =
\beta$ condition which corresponds to macroscopic Galilean
invariance is not satisfied generally. For more general
spin-dependent disorder models or more realistic exchange
splitting values $\alpha$ and $\beta$ are never equal.

These considerations to not apply directly to transition metal
ferromagnets because of $s-d$ hybridization and because of the
large $d$-orbital contribution to the minority-spin density of
states.  It is nevertheless true that the two reactive term
coefficients can be expressed approximately as the sum of $s$ and
$d$ orbital contributions.  In the absence of spin-orbit coupling,
the coefficient of $\partial \hat \Omega/\partial t$ in
Eq.~(\ref{eq:LLGwithSTTs}) is rigorously equal to one because the
Berry phase is proportional to the total spin-density, including
the $s$ and the dominant $d$ contribution. Similarly the reactive
coefficient of $\nabla \hat \Omega$ can be understood
\cite{nunez2006} in terms of the cancellation between convective
and precessional contributions to spin-dynamics in the static
limit. It follows that the $d$-orbital weight in this reactive
coefficient is not zero, as in the $s-d$ model, but still
relatively smaller than the $d$ contribution to the reactive
time-derivative coefficient. These considerations suggest that
$\beta/\alpha$ will tend to be larger than one in most transition
metal ferromagnets. The main challenges to addressing this issue
more quantitatively for a specific material are achieving an
understanding of the nature of its spin-independent and
spin-dependent disorder, accounting for the spin-orbit coupling
present in the bands of the perfect crystal, and evaluating the
vertex corrections (whose essential role is established by these
toy model calculations) in systems with complex band structures.

\begin{acknowledgments}
The authors gratefully acknowledge fruitful conversations with Ar.
Abanov, S. Barnes, G. Bauer, L. Keldysh, H. Kohno, P. Levy, A.
Rosch, W. Saslow, H. Stoof, G. Tatara, and Y. Tserkovnyak. This
work was partially supported by ONR under Grant No.\
ONR-N000140610122, by the NSF under Grants No.\ DMR-0547875 and
DMR-0606489, by the SRC-NRI (SWAN), and by the Welch foundation.
Jairo Sinova is a Cottrell Scholar of Research Corporation. ASN is
partially funded by Proyecto FSM0204
\end{acknowledgments}

\appendix* \section{Calculation of the retarded, imaginary and Keldysh
parts of the spin-density spin-density response function} \label{sec:appendix}
\begin{widetext}
The spin-density spin-density response function in
Eq.~(\ref{eq:respfctzerothplusvertex}) is given by
\begin{equation}
\label{eq:respfctsplitasasum}
  \Pi_{ab} (\bq;t,t') = \Pi^0_{ab} (\bq;t,t') + \Pi^1_{ab}
  (\bq;t,t')~,
\end{equation}
where the first term on the right-hand side is the lowest-order
diagram in Fig.~\ref{fig:spindensity}~(a) and the second term is
the vertex correction in Fig.~\ref{fig:spindensity}~(b). In the
first part of this appendix we evaluate the lowest-order diagram.
In the second part the vertex correction is calculated.
\subsection{No vertex corrections}
Without vertex corrections the response function
\begin{eqnarray}
\label{eq:respfctzeroth} \Pi^0_{ab} (\bq;t,t') = \frac{i
\Delta^2}{8 \hbar} \int \frac{d {\bf p}}{(2\pi)^3} {\rm Tr} \left[
\tau_a G ({\bf p}+{\bf q};t,t') \tau_{b} G({\bf p};t',t)
  \right]~,
\end{eqnarray}
is shown diagrammatically in Fig.~\ref{fig:spindensity}~(a). The
analytic pieces are now easily determined from the above result
and given by
\begin{eqnarray}
\label{eq:respfctzerothplusvertexgrlandless} && \Pi^{0,>}_{ab}
(\bq;t,t') = \frac{i \Delta^2}{8 \hbar}\rule{0mm}{7mm} \int
\frac{d {\bf p}}{(2\pi)^3} {\rm Tr} \left[ \tau_a G^> ({\bf
p}+{\bf q};t,t') \tau_{b} G^<({\bf p};t',t)
  \right]~; \nonumber \\
&& \Pi^{0,<}_{ab} (\bq;t,t') = \frac{i \Delta^2}{8
\hbar}\rule{0mm}{7mm} \int \frac{d {\bf p}}{(2\pi)^3} {\rm Tr}
\left[ \tau_a G^< ({\bf p}+{\bf q};t,t') \tau_{b} G^>({\bf
p};t',t)
  \right]~.
\end{eqnarray}
Using the results in Eqs.~(\ref{eq:gfkeldysh}-\ref{eq:gfretadv})
we have for the retarded and advanced components of the
zero-momentum Fourier-transformed response function that
\begin{eqnarray}
\label{eq:respfctretadvnovertex}
 \Pi^{0,(\pm)}_{ab} (\bq,\omega)
  =-\frac{\Delta^2}{8} \int\!\frac{d\epsilon}{(2\pi)}\int\!\frac{d\epsilon'}{(2\pi)}
  \int\!\frac{d\bp}{(2\pi)^3} \left[ \frac{N_{\rm F}(\epsilon-\mu) - N_{\rm F}(\epsilon'-\mu)}{\epsilon-\epsilon'-\hbar
  \omega^\pm}\right]
  {\rm Tr} \left[ \tau_a A (\bp,\epsilon) \tau_{b} A (\bp,\epsilon')
  \right]~.
\end{eqnarray}
Expanding for small energies we find that
\begin{eqnarray}
\label{eq:respfctretadvnovertexexpansion}
  && \Pi^{0,(\pm)}_{ab} (\bq,\omega)
 \simeq
  -\frac{\Delta^2}{8}  \int\!\frac{d\epsilon}{(2\pi)}\int\!\frac{d\epsilon'}{(2\pi)}
  \int\!\frac{d\bp}{(2\pi)^3} \left[N_{\rm F}(\epsilon-\mu) - N_{\rm F}(\epsilon'-\mu)\right] \frac{{\mathcal P}}{(\epsilon-\epsilon')}
  {\rm Tr} \left[ \tau_a A (\bp,\epsilon) \tau_{b} A (\bp,\epsilon')
  \right] \nonumber \\
&& -\frac{\Delta^2 \hbar \omega }{8}
\int\!\frac{d\epsilon}{(2\pi)}\int\!\frac{d\epsilon'}{(2\pi)}
  \int\!\frac{d\bp}{(2\pi)^3} \left[N_{\rm F}(\epsilon-\mu) - N_{\rm F}(\epsilon'-\mu)\right] \frac{{\mathcal P}}{(\epsilon-\epsilon')^2}
  {\rm Tr} \left[ \tau_a A (\bp,\epsilon) \tau_{b} A (\bp,\epsilon')
  \right] \nonumber \\
&& \mp \frac{i\Delta^2 \hbar \omega }{32\pi}
  \int\!\frac{d\bp}{(2\pi)^3}
  {\rm Tr} \left[ \tau_a A (\bp,\mu) \tau_{b} A (\bp,\mu)
  \right]~.
\end{eqnarray}
From this find $\Pi^{0,(\pm)}_{xx} (\bq,\omega)=
\Pi^{0,(\pm)}_{yy} (\bq,\omega)$, and $\Pi^{0,(\pm)}_{xy}
(\bq,\omega)=-\Pi^{0,(\pm)}_{yx} (\bq,\omega)$. Carrying out the
remaining integrations we have, in the limit $\gamma_\sigma/M \to
0$, that
\begin{eqnarray}
\label{eq:retadvresponsfctresultnovertex}
   \Pi^{0,(\pm)}_{xx} (\bq,\omega)
   &=& \frac{\Delta^2\rho_{\rm s}}{8 M} \pm \frac{i \Delta^2\pi \hbar \omega}{8} \left\{
   \frac{n_{\rm i} u^2 \nu_\uparrow \nu_\downarrow + n_{\rm s} u_{\rm s}^2
   \left[ \overline{S_\perp^2} \left(\nu_\uparrow^2 + \nu_\downarrow^2 \right)
   + \overline{S_z^2} \nu_\uparrow \nu_\downarrow \right]
   }{M^2}
   \right\}~, \nonumber \\
  \Pi^{0,(\pm)}_{xy} ({\bf 0},\omega)
    &=&  \frac{\Delta^2 \rho_{\rm s}}{16 M^2} i \hbar \omega~.
\end{eqnarray}

The Keldysh component of the response function is in first
instance given by
\begin{eqnarray}
\label{eq:keldyshcomponentrespfctnovertex}
   &&  \Pi^{0,{\rm K}}_{ab} ({\bf q},\omega)
  =  \frac{\pi i \Delta^2}{4} \int\!\frac{d\epsilon}{(2\pi)}\int\!\frac{d\epsilon'}{(2\pi)}
  \int\!\frac{d\bp}{(2\pi)^3} \delta (\hbar \omega-\epsilon+\epsilon')
  \nonumber \\
 && \times \left\{ \left[ 1-N_{\rm F}(\epsilon-\mu)\right] N_{\rm F}(\epsilon'-\mu)
  +N_{\rm F}(\epsilon-\mu) \left[ 1-N_{\rm F}(\epsilon'-\mu)\right]  \right\}
  {\rm Tr} \left[ \tau_a A (\bp,\epsilon) \tau_b A (\bp,\epsilon')
  \right]~.
\end{eqnarray}
From this we see that $\Pi^{0,{\rm K}}_{xx} ({\bf q},\omega)=
\Pi^{0,{\rm K}}_{yy} ({\bf q},\omega)$, and $ \Pi^{0,{\rm K}}_{xy}
(\bq,\omega)= \Pi^{0,{\rm K}}_{yx} (\bq,\omega)=0$. We find that
\begin{equation}
\label{eq:resultkeldyshnovertex}
    \Pi^{0,{\rm K}}_{xx} ({\bf q} ,\omega)
    = - \frac{ i \pi \Delta^2  k_{\rm B} T}{2}  \left\{
   \frac{n_{\rm i} u^2 \nu_\uparrow \nu_\downarrow + n_{\rm s} u_{\rm s}^2
   \left[ \overline{S_\perp^2} \left(\nu_\uparrow^2 + \nu_\downarrow^2 \right)
   + \overline{S_z^2} \nu_\uparrow \nu_\downarrow \right]
   }{M^2}\right\}~.
\end{equation}
Moreover, we have that
\begin{equation}
\label{eq:fdtheoremnovertexcorrs}
   \Pi^{0,{\rm K}}_{xx} (\bq ,\omega) = \pm 2 i \left[ 2 N_{\rm B} (\hbar \omega) +1
\right] {\rm Im}  \Pi^{0,({\pm})}_{xx} (\bq, \omega)~,
\end{equation}
where $N_{\rm B} (x)$ is the Bose distribution function. This is
the fluctuation-dissipation theorem, which emerges naturally from
the formalism.

\subsection{Vertex correction}
The first-order vertex correction is shown in
Fig.~\ref{fig:spindensity}~(b), and given by
\begin{eqnarray}
\label{eq:respkeldysh1stvertex}
 && \Pi^{1}_{ab} (\bq;t,t') =  \frac{i \Delta^2}{8 \hbar^3} \sum_{a' \in
  \{0,x,y,z\}}\sigma_{a'} \int\! \frac{d {\bf p}_1}{(2\pi)^3}
  \int\! \frac{d {\bf p}_2}{(2\pi)^3}
  \int_{{\mathcal C}^\infty}\!dt''\int_{{\mathcal C}^\infty}dt''' \nonumber \\
  && \times
  {\rm Tr} \left[ \tau_a G (\bq+\bp_1;t,t''') \tau_{a'}
  G (\bq+\bp_2;t''',t')\tau_{b} G (\bp_2;t',t'')\tau_{a'} G (\bp_1;t'',t)
  \right]~.
\end{eqnarray}

Before we proceed, we state some rules for calculus involving the
Keldysh contour. From an equation like
\begin{equation}
 A(t,t') = \int_{{\mathcal C}^\infty} dt'' B (t,t'') C(t'',t')~,
\end{equation}
we find the analytic pieces as
\begin{equation}
 A^{\gtrless}(t,t') =
  \int_{-\infty}^\infty dt'' B^{(+)} (t,t'') C^{\gtrless} (t'',t')
  +\int_{-\infty}^\infty dt'' B^{\gtrless} (t,t'') C^{(-)} (t'',t')~,
\end{equation}
where the retarded and advanced components are defined in
Eq.~(\ref{eq:defretadv}). It is then also straightforward to show
that
\begin{equation}
A^{(\pm)} (t,t') = \int_{-\infty}^\infty dt'' B^{(\pm)} (t,t'')
C^{(\pm)} (t'',t')~.
\end{equation}

Using these rules we find from Eq.~(\ref{eq:respkeldysh1stvertex})
\begin{eqnarray}
\label{eq:retadvfull1stvertex}
  &&  \Pi^{1, (\pm)}_{ab} (\bq,\omega)
   = - \frac{\Delta^2}{8}\sum_{{a'} \in \{0,x,y,z\}} \sigma_{a'} \int\!\frac{d\epsilon}{(2\pi)}
  \int\! \frac{d\epsilon'}{(2\pi)}
   \int \frac{d\bp_1}{(2\pi)^3} \int \frac{d\bp_2}{(2\pi)^3}
   \frac{1}{\hbar \omega^\pm-\epsilon+\epsilon'}
   \nonumber \\
   && \times {\rm Tr} \left\{
   \left[
  \tau_a G^{(+)} (\bq+\bp_1;\epsilon) \tau_{a'}
  G^> (\bq+\bp_2;\epsilon) + \tau_a G^> (\bq+\bp_1;\epsilon) \tau_{a'}
  G^{(-)} (\bq+\bp_2;\epsilon)\right] \right.\nonumber \\ && \left.\left[\tau_{b} G^{(+)} (\bp_2;\epsilon')\tau_{a'}
  G^<
  (\bp_1;\epsilon')+\tau_{b} G^< (\bp_2;\epsilon')\tau_{a'}
  G^{(-)}
  (\bp_1;\epsilon')\right]
   \right. \nonumber \\
   && - \left. \left[
  \tau_a G^{(+)} (\bq+\bp_1;\epsilon) \tau_{a'}
  G^< (\bq+\bp_2;\epsilon) + \tau_a G^< (\bq+\bp_1;\epsilon) \tau_{a'}
  G^{(-)} (\bq+\bp_2;\epsilon)\right] \right.\nonumber \\ && \left.\left[\tau_{b} G^{(+)} (\bp_2;\epsilon')\tau_{a'}
  G^>
  (\bp_1;\epsilon')+\tau_{b} G^> (\bp_2;\epsilon')\tau_{a'}
  G^{(-)}
  (\bp_1;\epsilon')\right] \right\}~.
\end{eqnarray}
The Keldysh component reads
\begin{eqnarray}
\label{eq:keldyshfull1stvertex}
  &&  \Pi^{1, {\rm K}}_{ab} (\bq,\omega)
   = \frac{i \pi \Delta^2}{4} \sum_{{a'} \in \{0,x,y,z\}} \sigma_{a'} \int\!\frac{d\epsilon}{(2\pi)}
  \int\! \frac{d\epsilon'}{(2\pi)}
   \int \frac{d\bp_1}{(2\pi)^3} \int \frac{d\bp_2}{(2\pi)^3}
  \delta(\hbar \omega-\epsilon+\epsilon')
   \nonumber \\
   && \times {\rm Tr} \left\{
   \left[
  \tau_a G^{(+)} (\bq+\bp_1;\epsilon) \tau_{a'}
  G^> (\bq+\bp_2;\epsilon) + \tau_a G^> (\bq+\bp_1;\epsilon) \tau_{a'}
  G^{(-)} (\bq+\bp_2;\epsilon)\right] \right.\nonumber \\ && \left.\left[\tau_{b} G^{(+)} (\bp_2;\epsilon')\tau_{a'}
  G^<
  (\bp_1;\epsilon')+\tau_{b} G^< (\bp_2;\epsilon')\tau_{a'}
  G^{(-)}
  (\bp_1;\epsilon')\right]
   \right. \nonumber \\
   && + \left. \left[
  \tau_a G^{(+)} (\bq+\bp_1;\epsilon) \tau_{a'}
  G^< (\bq+\bp_2;\epsilon) + \tau_a G^< (\bq+\bp_1;\epsilon) \tau_{a'}
  G^{(-)} (\bq+\bp_2;\epsilon)\right] \right.\nonumber \\ && \left.\left[\tau_{b} G^{(+)} (\bp_2;\epsilon')\tau_{a'}
  G^>
  (\bp_1;\epsilon')+\tau_{b} G^> (\bp_2;\epsilon')\tau_{a'}
  G^{(-)}
  (\bp_1;\epsilon')\right] \right\}~.
\end{eqnarray}
We focus now on the imaginary part of $ \Pi^{1,(\pm)}_{xx}
(\bq,\omega) = \Pi^{1,(\pm)}_{yy} (\bq,\omega)$ since these
determine the Gilbert damping constant. Carrying out the momentum
and energy integrals results in
\begin{equation}
\label{eq:retadv1stvertexlowenergy}
   \Pi^{1,(\pm)}_{xx} (\bq, \omega)
   = \mp \frac{i \pi \Delta^2 \hbar \omega}{8} \left[
   \frac{n_{\rm i} u^2 \nu_\uparrow \nu_\downarrow - n_{\rm s} u_{\rm s}^2
   \overline{S_z^2} \nu_\uparrow \nu_\downarrow
   }{M^2}
   \right]~.
\end{equation}
The corresponding Keldysh part is given by
\begin{equation}
\label{eq:keldysh1stvertexlowenergy}
   \Pi^{1,{\rm K}}_{xx} (\bq,\omega)
   =\frac{\pi i \Delta^2 k_{\rm B} T}{2} \left[
   \frac{n_{\rm i} u^2 \nu_\uparrow \nu_\downarrow - n_{\rm s} u_{\rm s}^2
   \overline{S_z^2} \nu_\uparrow \nu_\downarrow
   }{M^2}
   \right]~.
\end{equation}
Adding the results from
Eqs.~(\ref{eq:retadvresponsfctresultnovertex})~and~(\ref{eq:retadv1stvertexlowenergy})
we get the result presented in
Eq.~(\ref{eq:retadvresponsfctfinalresult}). Similarly, adding the
result in Eq.~(\ref{eq:resultkeldyshnovertex}) to
Eq.~(\ref{eq:keldysh1stvertexlowenergy}) reproduces
Eq.~(\ref{eq:keldyshtotallowenergy}).

\end{widetext}


\begin{thebibliography}{99}
\bibitem{note1} Note that in the long-wavelength limit the coefficients of the spin transfer
torques and the Gilbert damping parameter can not depend on position.
\end{thebibliography}
\end{document}